\documentclass[twocolumn,prl,superscriptaddress]{revtex4-1}

\usepackage{amsmath,amssymb,amsfonts,mathrsfs}
\usepackage{amsthm}
\usepackage{graphicx}
\usepackage{dcolumn}
\usepackage{bm}
\usepackage{color}

\begin{document}

\title{Experimental Realization of Universal Time-optimal non-Abelian Geometric Gates}

\author{Zhikun Han}
\thanks{These authors contributed equally to this work.}
\affiliation{National Laboratory of Solid State Microstructures, School of Physics, Nanjing University, Nanjing 210093, China}
\author{Yuqian Dong}
\thanks{These authors contributed equally to this work.}
\affiliation{National Laboratory of Solid State Microstructures, School of Physics, Nanjing University, Nanjing 210093, China}
\author{Baojie Liu}
\thanks{These authors contributed equally to this work.}
\affiliation{Institute for Quantum Science and Engineering, and Department of Physics, Southern University of Science and Technology, Shenzhen 518055, China}
\author{Xiaopei Yang}
\affiliation{National Laboratory of Solid State Microstructures, School of Physics, Nanjing University, Nanjing 210093, China}
\author{Shuqing Song}
\affiliation{National Laboratory of Solid State Microstructures, School of Physics, Nanjing University, Nanjing 210093, China}
\author{Luqing Qiu}
\affiliation{National Laboratory of Solid State Microstructures, School of Physics, Nanjing University, Nanjing 210093, China}
\author{Danyu Li}
\affiliation{National Laboratory of Solid State Microstructures, School of Physics, Nanjing University, Nanjing 210093, China}
\author{Ji Chu}
\affiliation{National Laboratory of Solid State Microstructures, School of Physics, Nanjing University, Nanjing 210093, China}
\author{Wen Zheng}
\affiliation{National Laboratory of Solid State Microstructures, School of Physics, Nanjing University, Nanjing 210093, China}
\author{Jianwen Xu}
\affiliation{National Laboratory of Solid State Microstructures, School of Physics, Nanjing University, Nanjing 210093, China}
\author{Tianqi Huang}
\affiliation{National Laboratory of Solid State Microstructures, School of Physics, Nanjing University, Nanjing 210093, China}

\author{Zhimin Wang}
\affiliation{National Laboratory of Solid State Microstructures, School of Physics, Nanjing University, Nanjing 210093, China}
\author{Xiangmin Yu}
\affiliation{National Laboratory of Solid State Microstructures, School of Physics, Nanjing University, Nanjing 210093, China}
\author{Xinsheng Tan}
\email{tanxs@nju.edu.cn}
\affiliation{National Laboratory of Solid State Microstructures, School of Physics, Nanjing University, Nanjing 210093, China}
\author{Dong Lan}
\affiliation{National Laboratory of Solid State Microstructures, School of Physics, Nanjing University, Nanjing 210093, China}
\author{Man-Hong Yung}
\email{yung@sustech.edu.cn}
\affiliation{Institute for Quantum Science and Engineering, and Department of Physics, Southern University of Science and Technology, Shenzhen 518055, China}
\author{Yang Yu}
\affiliation{National Laboratory of Solid State Microstructures, School of Physics, Nanjing University, Nanjing 210093, China}

\begin{abstract}

Based on the geometrical nature of quantum phases, non-adiabatic holonomic quantum control (NHQC) has become a standard technique for enhancing robustness in constructing quantum gates. However, the conventional approach of NHQC is sensitive to control instability, as it requires the driving pulses to cover a fixed pulse area. Furthermore, even for small-angle rotations, all operations need to be completed with the same duration of time. Here we experimentally demonstrate a time-optimal and unconventional approach of NHQC (called TOUNHQC), which can optimize the operation time of any holonomic gate. Compared with the conventional approach, TOUNHQC provides an extra layer of robustness to decoherence and control errors.  The experiment involves a scalable architecture of superconducting circuit, where we achieved a fidelity of $99.51\%$ for a single qubit gate using interleaved randomized benchmarking. Moreover, a two-qubit holonomic control-phase gate has been implemented where the gate error can be reduced by as much as $18\%$ compared with NHQC. 

\end{abstract}

\maketitle
\emph{Introduction}.\textbf{--} Quantum computation \cite{Nielson2002,Arute2019}, which can provide an unprecedented computational power over classical computation, relies heavily on high fidelity quantum manipulations. However, there are two main obstacles to achieve high fidelity quantum gates. One is the decoherence caused by inevitable interaction between environment and quantum systems, and the other is the imperfection of control pulses caused by the crosstalk between control lines and signal deformation. To cope with these problems, different approaches have been developed, including quantum error correction (QEC) \cite{Gottesman1997,Knill2005} and decoherence-free systems (DFS) \cite{Duan1997,Zanardi1997,Lidar1998}. Recently, it has been proposed to utilize geometric phases \cite{Wilczek1984,Berry1984,Aharonov1987} in quantum system to construct gate operation, for constructing geometric gates due to their inherent robustness under fluctuations of the orbital trajectory in control space  \cite{Chiara2003,Zhusl2005,Fliipp2009,Berger2013}. 

Geometric phase, as the key element in this protocol, has been studied theoretically and experimentally for decades \cite{Falci2000,Whitney2005,Leek2007,Mottonen2008}: geometric phase can be either Abelian (Berry phase) or non-Abelian. The non-abelian geometric phase, due to its noncommutativity, can naturally lead to universal quantum computation (so-called Holonomic quantum computation) \cite{Zanardi1999,Pachos2000,Duan2001,Albert2016}. Originally, Holonomic quantum computation (HQC) was constructed via adiabatic evolutions, which lead to lengthy gate time and hence induce higher decoherence error. Subsequently, Nonadiabatic Holonomic quantum computation (NHQC)\cite{Sjoqvist2012,Xu2015,zhu2019,Herterich2016}, in which the adiabatic constrain does not exist hence shorten the gate time, was proposed and several experimental demonstrations of NHQC have been realized on NMR \cite{Feng2013,Li2017}, NV center in diamond \cite{Arroyo2014,ZuC2014,Sekiguchi2017,Zhou2017,Nagata2018,Ishida2018} and superconducting circuits \cite{Abdumalikov2013,Xu2018,Egger2019}.  However, this NHQC condition imposes stringent requirements on the driving Hamiltonian; the systematic errors would introduce additional fluctuating phase shifts, smearing the geometric phases \cite{Zheng2016,Yan2019,Liu2019}.  Furthermore, even for small-angle rotations, all operations need to be completed with the same duration of time.

Here, we experimentally demonstrate a new scheme of holonomic quantum gates, where a unconventional nonadiabatic and non-Abelian geometric phase is employed to construct quantum gates. The scheme is called time-optimal unconventional non-adiabatic holonomic quantum computation (TOUNHQC) \cite{liu2020brachistochronic}, which can construct arbitrary nonadiabatic and non-Abelian geometric gates with the minimal time. Thus, TOUNHQC offers the unique stability of robustness to decoherence errors and control parameter variations. 
     
For an experimental demonstration, we experimentally realize a universal set of single-qubit and two-qubit unconventional holonomic quantum gates in an Xmon-type of superconducting circuit. Compared with the conventional NHQC implementations, this scheme is more robust against control error by removing the fixed pulse area limitation. Moreover, by combining with time-optimal technology, we further minimize the pulse time for both single and two qubit gate, hence achieving less decoherence error than conventional NHQC. We use gate in single qubit and C-phase gate in two qubits as examples to demonstrate the advantage of our routine.

\emph{Setting the stage.}\textbf{--} Similar to the traditional holonomic gate, an $\Lambda$ shape~\cite{Abdumalikov2013,Gu2017}  Hamiltonian is constructed in our quantum system by applying driving microwave, which can be written as 
\begin{equation}
H_{1}(t)=\frac{\Omega_{0e}(t)}{2}e^{i\phi_{0}}|0\rangle \langle{e}|+
\frac{\Omega_{1e}(t)}{2}e^{i\phi_{1}}|1\rangle \langle{e}|+H.c.,
\label{H1}
\end{equation}
where energy levels $|0\rangle$ and $|1\rangle$ are computational basis which are both coupled to an auxiliary level $|e\rangle$. By setting $\Omega_{0e}(t)=\Omega(t) \sin(\theta/2)$ and $\Omega_{1e}(t)=\Omega(t) \cos(\theta/2)$, we can obtain bright state $|b\rangle=\sin(\theta/2) e^{i\phi}|0\rangle+\cos(\theta/2)|1\rangle$ and dark state $|d\rangle=\cos(\theta/2) e^{i\phi}|0\rangle-\sin(\theta/2)|1\rangle$, where $\phi\equiv\phi_{0}-\phi_{1}$. We shall keep $\theta$ time-independent, the above Hamiltonian Eq.~(\ref{H1}) can then be rewritten as:
\begin{equation}
H_{1}(t)=\frac{\Omega(t)}{2}e^{i\phi_{1}(t)}|b\rangle \langle{e}|+H.c .
\label{H2}
\end{equation}

\emph{Time-optimal unconventional holonomic gates.}\textbf{--} To realize holonomic gates, we need to
choose a set of auxiliary states $\{|\mu_{k}(t)\rangle\}$, which satisfy the following conditions of (i) the cyclic evolution $\Pi_{1}(0)=\Pi_{1}(\tau)=|d\rangle\langle d|$, $\Pi_{2}(0)=\Pi_{2}(\tau)=|b\rangle\langle b|$, and (ii) the von Neumann equation $\frac{d}{d t} \Pi_{k}(t)=-i\left[H_{1}(t), \Pi_{k}(t)\right]$,  where $\Pi_{k}(t) \equiv\left|\mu_{k}(t)\right\rangle\left\langle\mu_{k}(t)\right|$ denotes the projector of the auxiliary basis. Here, we choose a decoupled dark state $|\mu_{1}\rangle=|d\rangle$ and an orthogonal state $|\mu_{2}(t)\rangle$, which can be parameterized with two angles $\chi$, $\alpha(t)$ and $\eta(t)$ as $\left|\mu_{2}(t)\right\rangle=(\cos \eta(t)-i\sin\eta(t)\cos\chi)|b\rangle-i\sin\eta(t)\sin\chi e^{i \alpha(t)}|e\rangle$. Using the von Neumann equation, we obtain the following coupled differential equations,
\begin{equation}\label{CDE}
\Omega(t)=-\dot{\alpha} (t) \tan \chi, \quad \dot{\phi_{1}}(t)=\dot{\alpha}(t)=-2\dot{\eta}(t)\cos\chi
\end{equation}
In fact, we have many choices to pick the variables $\eta(t)$, $\alpha(t)$ as longle as they meet the coupled equation and the cyclic evolution condition $\eta(0)=0$ and $\eta(\tau)=\pi$.

\begin{figure}[bt]
	\includegraphics[width=8.5cm,height=4.25cm]{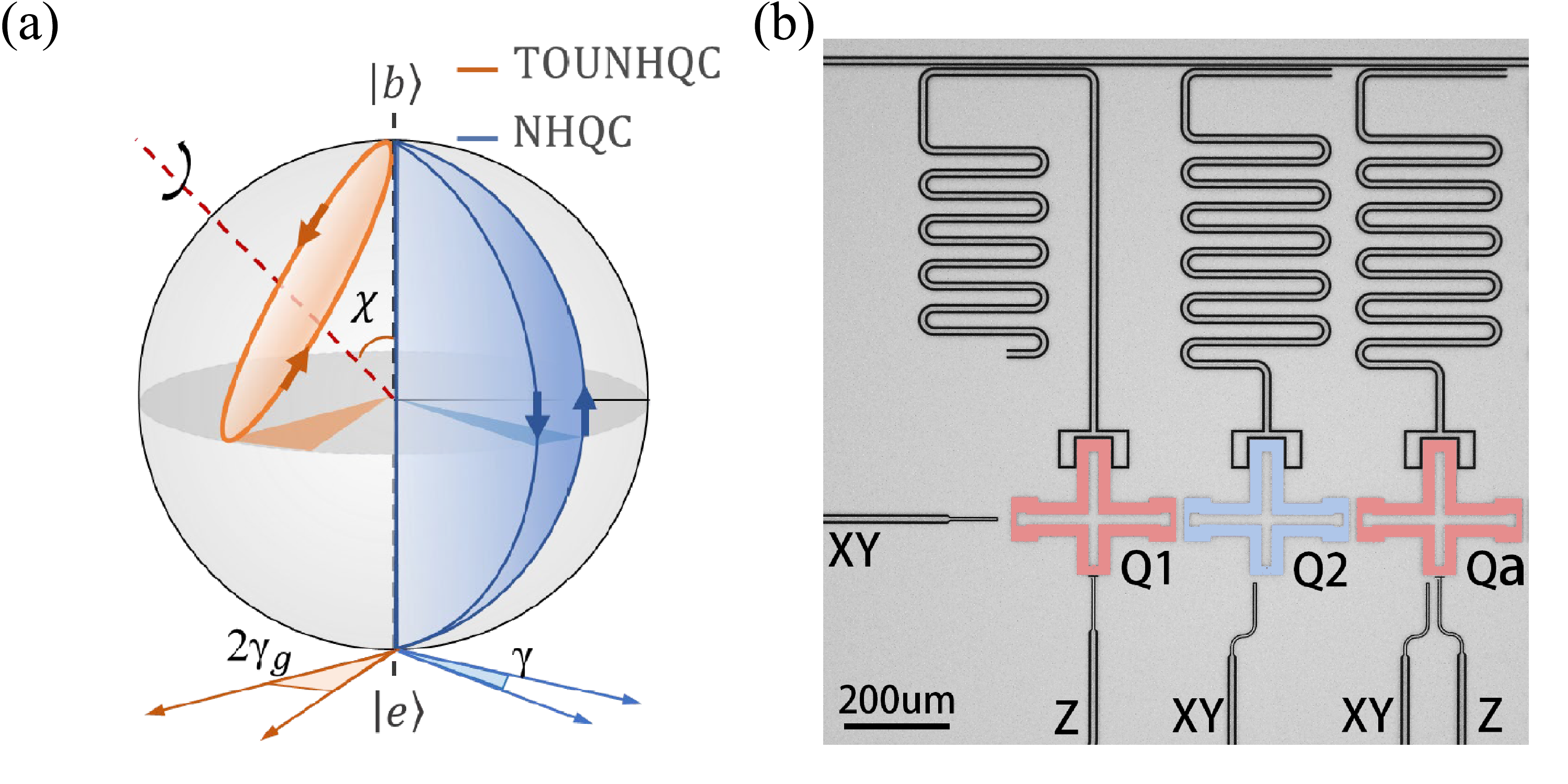}
	\caption{\label{fig:epsart} (a) A geometric picture of the single-qubit Holonomic gates in Bloch sphere representation. The orange line is the evolution trajectory of TOUNHQC protocol while the blue one illustrates conventional NHQC protocol. $\gamma_{g}$ and $\gamma$ are geometric phases acquired from these two gate operations respectively. (b) Optical image of the six qubits sample (three not shown) used in our experiments.
	}
\end{figure}

Now, we demonstrate how to build up universal arbitrary holonomic single-qubit gates. Let us start with the an arbitrary initial state as $|\psi(0)\rangle=a_{1}|\mu_{1}\rangle+a_{2}|\mu_{2}(0)\rangle$ with $a_{1,2}=\left\langle\mu_{1,2}(0)|\psi(0)\right\rangle$. Under a cyclic evolution, the state $|\mu_{2}\rangle$ gains a unconventional geometric phase~\cite{liu2020brachistochronic,Zhusl2003,du2006experimental}  that is $|\mu_{2}(\tau)\rangle=e^{-i(\gamma_{g}+\gamma_{d})}|\mu_{2}(0)\rangle=e^{-i\phi_{g}}|\mu_{2}(0)\rangle$ corresponding to the geometric phase $\gamma_{g}=\int^{\alpha(\tau)}_{\alpha(0)}\sin^{2}\eta(t)\sin^{2}\chi d\alpha$ and the dynamical phase $\gamma_{d}=-\gamma_{g}+\phi_{g}$. Note that the unconventional  geometric phase $\gamma\equiv\gamma_{g}=\alpha(\tau)-\alpha(0)$ is just the half of the azimuthal angle difference between initial and final points, which is also only determined by the geometric feature of the evolution path. As a result, the final time evolution operator on the subspace $\{|b\rangle,|d\rangle\}$ is given by $U(\tau)=e^{i \gamma}|b\rangle\langle b|+| d\rangle\langle d|$. 

Consequently,  the unconventional non-Abelian geometric gate can be spanned by the logical $\{|0\rangle, |1\rangle\}$ basis as,
\begin{equation}
\begin{aligned} U(\theta, \phi, \gamma) &=\left(\begin{array}{cc}c_{\frac{\gamma}{2}}-i s_{\frac{\gamma}{2}} c_{\theta} & -i s_{ \frac{\gamma}{2}} s_{\theta} e^{i \phi_{1}}  \\ -i s_{ \frac{\gamma}{2}} s_{\theta} e^{-i \phi} & c_{\frac{\gamma}{2}}+i s_{\frac{\gamma}{2}} c_{\theta} \end{array}\right) \\ &=\exp \left(-i \frac{\gamma}{2} \mathbf{n} \cdot \sigma\right) \end{aligned} \ ,
\label{U}
\end{equation}
where ${c_x} \equiv \cos x$ and ${s_x} \equiv \sin x$. 
Note this operation corresponds to an arbitrary rotation around the axis $\textbf{n}=(\sin{\theta}\cos{\phi},\sin{\theta}\sin{\phi},\cos{\theta})$ by an angle of $\gamma$, which picks up a global phase $\gamma/2$. Therefore, it is feasible to implement the unconventional holonomic gate by designing particular $\phi$ and $\gamma$ .
 
To further speed up the nonadiabatic holonomic gates, we combined our unconventional scheme with time-optimal control technology~\cite{Carlini2012,Wang2015} to \emph{shorten} the loop path and \emph{minimum} the gate time against the influence of the environmental induced decoherence effect. In the time-optimal scheme, the phase $\phi_{1}(t)$ in the Eq.~\ref{H2} equals to $2(\pi-\gamma)t/\tau$, with operation time $\tau=2\sqrt{\pi^2-(\pi-\gamma)^2}/\Omega_{0}$. Comparing with the conventional NHQC, the TOUNHQC scheme generally has a shorter gate time~\cite{liu2020brachistochronic}.

\emph{Experimental results of time-optimal holonomic single-qubit gate}\textbf{--} These experiments are demonstrated on a chain of superconducting Xmon-shaped qubits \cite{Clarke2008,Devoret2013,Barends2013}, as shown in Fig. 1(b). There are six qubits on the chip (three of them not shown) and each qubit is individually coupled to a $\lambda/4$ resonator, which is coupled to a common feed line for readout\cite{aSupp}. These qubits are coupled to each other with a static capacitive coupling strength of $g/2\pi\approx25$ MHz. Among them, two qubits highlighted in red (denoted as $Q_1$ and $Q_a$) are tunable transmons while the one in blue (denoted as $Q_2$) is a fixed-frequency transmon.

We first demonstrate single-qubit time-optimal holonomic gates by following the single-loop protocol in the reference. The experiment is performed on the three lowest energy levels $|0\rangle, |e\rangle, |1\rangle$ of a superconducting qubit ($Q_2$), as shown in Fig. 2(a). Among them, the ground state $|0\rangle$ and the second excited state $|1\rangle$ are used to construct the qubit computational basis, while the first excited state $|e\rangle$ is regarded as an auxiliary energy level. 

\begin{figure}[hbt]
\includegraphics[width=8.5cm,height=9cm]{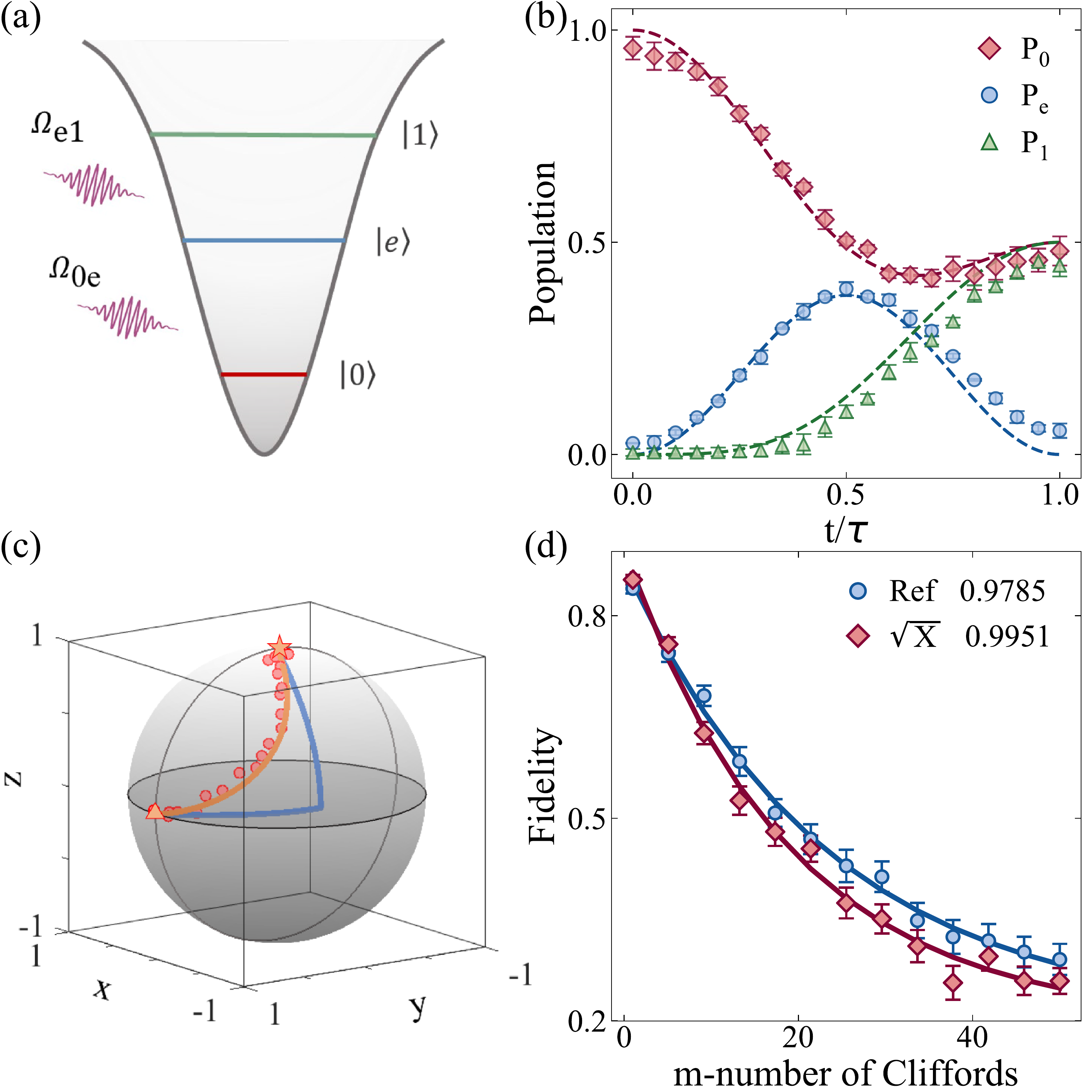}
\caption{\label{fig:epsart} (a) The energy level configuration of a qutrit. Two microwave are driven resonantly with $|0\rangle\leftrightarrow|e\rangle$ and $|e\rangle\leftrightarrow|1\rangle$ transitions. (b) States transfer during the $\sqrt{X}$ gate. The experimental data (blue circles, red diamonds and green triangles) shows a good agreement with the simulation results. (c) State evolution of $\sqrt{X}$ gate of TO-UNHQC demonstrated in the $|0\rangle-|1\rangle$ subspace, where the pentagram and triangle represent the beginning and end of the evolution procedure respectively. Solid dots and lines represent experimental data and theoretical simulation respectively. Theoretical simulation of the gate based on NHQC protocol is also shown for comparison.  (d) To evaluate the performance of our routine, we choose Clifford-based randomized benchmarking (RB) \cite{Chow2009,Knill2008,Magesan2011} to extract the fidelity.}
\end{figure}

Here we take a $\sqrt{X}$ gate as an example, which is performed on the logical qubit initialized in state $|0\rangle$. The rotation operator in Eq.~\ref{U} is set as $U(\pi/2,0,\pi/2)$. In our approach, we fix $\Omega_0/2\pi=8.660$ MHz, making the driving amplitudes $\Omega_{0e}/2\pi= \Omega_{e1}/2\pi=6.124$ MHz, with the evolution time $\tau=100$ ns. To characterize state evolution in the gate operation, we execute state tomography during entire procedure. The populations of $|0\rangle$, $|1\rangle$ and $|e\rangle$ are illustrated in Fig.~2(b), while state trajectory in Bloch sphere spanned by computational basis is shown in Fig.~2(c).   There is a good agreement between the experimental data and the theoretical result. Furthermore,  we show the direct comparison of TOUNHQC (red) and NHQC (blue) in terms of the $\sqrt{X}$ gate. Notice that in our approach, the evolution trajectory in the Bloch sphere is shorter due to time-optimal control.  The total procedure times for the protocols are  $\tau_{TOUNHQC}=100$ ns and $\tau_{NHQC}=115.47$ ns respectively, displaying advantages of our protocol in operation time.

\begin{figure}[bt]
\includegraphics[width=8.5cm,height=10.625cm]{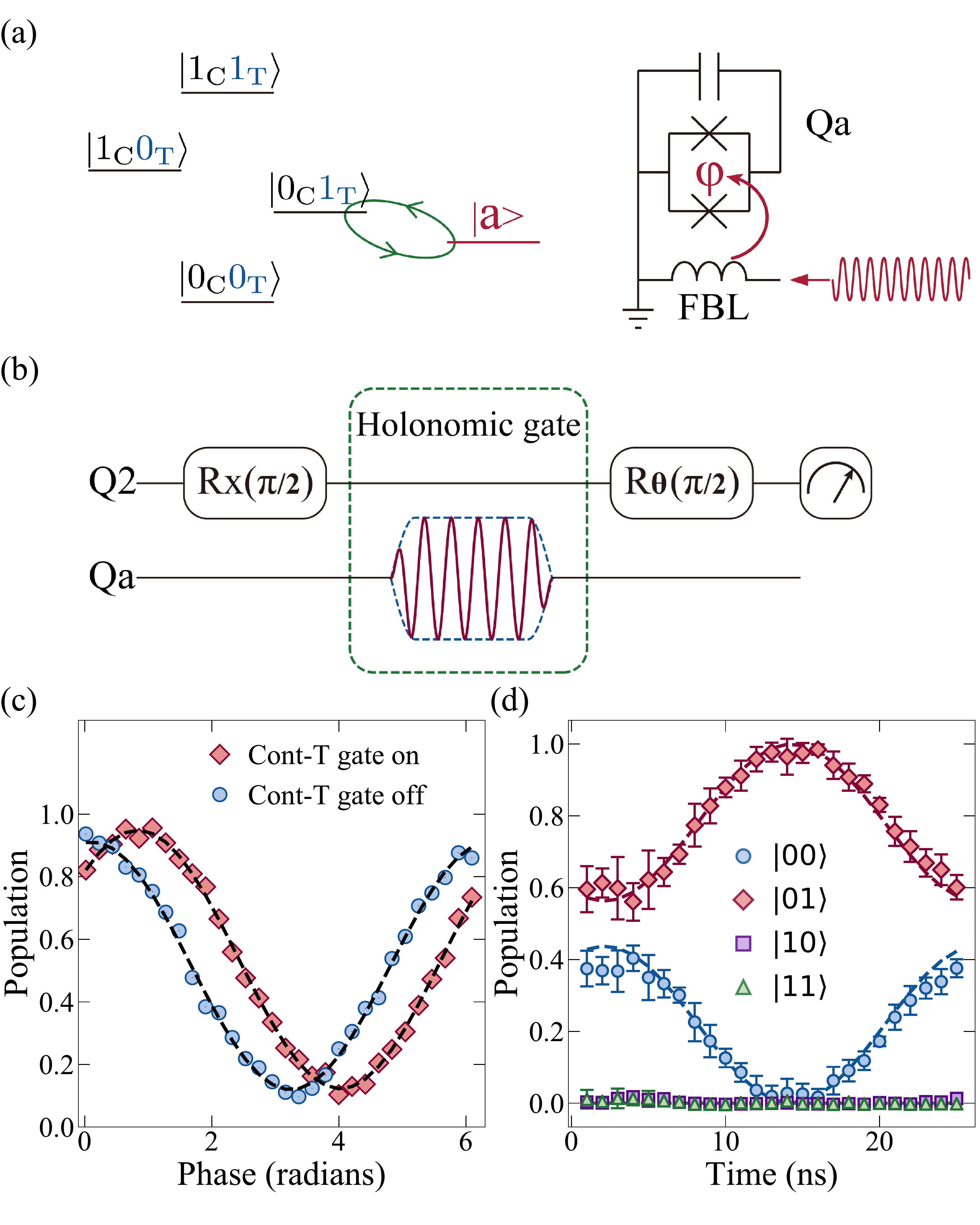}
\caption{\label{fig:epsart} (a)$Q_1$ (control qubit) and $Q_2$ (target qubit), form the Hilbert space spanned by $\{|\rm0_C0_T\rangle, |0_C1_T\rangle, |1_C0_T\rangle, |1_C1_T\rangle\}$. The two-qubit gates are realized by interaction between the $|01\rangle$ and an ancillary state $|a\rangle$ of $Q_a$. The interaction is activated by modulating the frequency of $Q_a$ with an ac flux through the flux bias line (FBL). (b) The pulse sequence to perform Ramsey interferometer. (c) The results of Ramsey interferometer with control-T gate on (red circle) or off (blue triangle). The dash lines are used as guidelines. (d) States transfer during the control-T gate. Experimental results fit well with numerical simulations (dash lines). The error bars indicate standard deviation of the data. }
\end{figure}

\emph{Experimental results of time-optimal holonomic two-qubit entangled gate.}\textbf{--} A universal set of gates requires a kind of two qubit operation. We can demonstrate nontrivial two-qubit C-phase gates of TOUNHQC using the three qubits sample. In practice, we choose control-$R_k$ gate as an example,
\begin{equation}
R_{k}=\begin{pmatrix} 1 & 0 \\ 0 & e^{i\frac{2\pi}{2^k}} \end{pmatrix},
\label{Rk}
\end{equation}
which is important for realizing a fast quantum Fourier transform \cite{Nielson2002}. Here we utilize the lowest two levels of $Q_1$ and $Q_2$ to form the logical computation space spanned by $\{|00\rangle, |01\rangle, |10\rangle, |11\rangle\}$ while use the first excited state of $Q_a$ (denoted as $|a\rangle$) as an ancilla state, as seen in the left panel of Fig.~3a. In the C-phase routine, $|b\rangle$ and $|e\rangle$ in Eq.~\ref{H2} are regarded as $|01\rangle$ and $|a\rangle$ respectively.  To implement C-phase gate,  we use parametric modulation to activate the interaction between $|01\rangle$ and $|a\rangle$ \cite{Caldwell2018,Didier2018,Reagor2018}. 

 Without loss of generality, we demonstrate the Control-T in practice, by setting value of $n$ in Eq.~\ref{Rk} as 3. The parameters $\theta,\phi,\gamma$ in Eq.\ref{U} are set as $\{0,0,\pi/4\}$ respectively\cite{Supp}. Consequently, we set $\xi=\pi/4$ in the experiment, then  the unitary operator in the subspace $\{|00\rangle, |01\rangle, |10\rangle, |11\rangle\}$ is written as
\begin{equation}
U=\begin{pmatrix} 1 & 0 & 0 & 0\\ 0 & e^{i\frac{\pi}{4}} & 0 & 0\\ 0 & 0 & 1 & 0\\ 0 & 0 & 0 & 1 \end{pmatrix}
\end{equation}

\begin{figure}[bt]
\includegraphics[width=8.5cm,height=12.75cm]{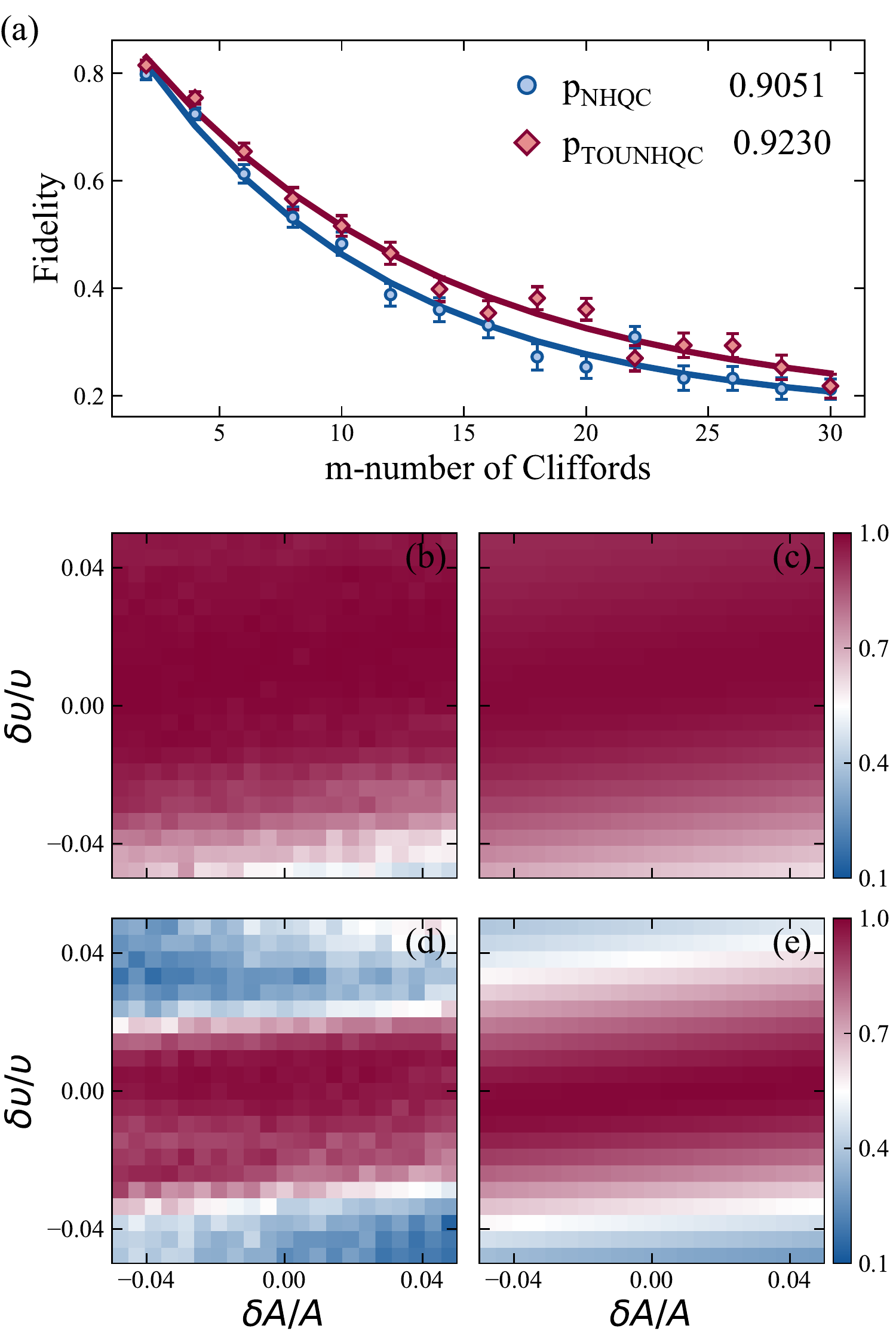}
\caption{\label{fig:epsart}(a) Superiority of the TOUNHQC scheme in terms of gate operation time. The Fidelity of the gate based on TOUNHQC (b) and NHQC (d) scheme varied as a function of both pulse amplitude error $\delta A/A$ and the pulse detuning error $\delta \upsilon/\upsilon$. Both the amplitude and detuning error are range from -0.05 to 0.05. (c) and (e) are the corresponding simulation results.
}
\end{figure}

To verify that $Q_2$ indeed picks up $\xi=\pi/4$ after the gate, we execute the Ramsey interferometer, whose pulse sequence is shown in Fig. 3(b). We initially prepare $Q_2$ in $(|0\rangle-i|1\rangle)/\sqrt{2}$ while $Q_1$ and $Q_a$ remain in $|0\rangle$. Then a $\pi/2$ pulse around axis ($\rm cos\rm \theta,\rm sin\rm \theta,0$) is applied right after the two-qubits Holonomic gate. As seen in Fig. 3(c), the final population with and without the Holonomic gate are plotted as the function of $\theta$ respectively, demonstrating the qubit indeed acquires a geometric phase about $\pi/4$. We also measured the populations of each states in the entire procedure besides the finite rising and falling edge, as shown in Fig. 3(d). It is noticed that our data is in good agreement with the simulation result \cite{Johansson2012}.

\emph{ The superiority of TOUNHQC.}\textbf{--} Next, we demonstrate the superiority of the TOUNHQC protocol in terms of gate operation time. Here we use the interleaved RB to calibrate the performance of time-optimal and conventional methods affected by decoherence. The pulse sequence of RB approach is set as: a randomly selected Clifford gate and a subsequent unit gate with the specific Holonomic T gate length. To compare these two approaches,  we set the same amplitude of driving microwave $\Omega$, resulting the gate times $\tau_{TOUNHQC}=\sqrt{7}\pi/2\Omega$ and $\tau_{NHQC}=2\pi/\Omega$. The shorter operation time means less decoherence effect. Therefore, RB with our time-optimal protocol has higher fidelity. As shown in Fig.~4(a), by fitting the curve of experimental
data with function $F=Ap^{m}+B$,  we obtain the experimental results for $P_{TOUNHQC}$ and $P_{NHQC}$ as 0.9230 and 0.9051 respectively. 

To further demonstrate that our scheme is more robust than the conventional NHQC, we measure gate fidelities under the imperfections of control parameters for both our unconventional Holonomic gate and the conventional Holonomic gate. The unattenuated fidelities here are defined by $F_{unatt}=\rm{Tr}(\rho_{th}\rho_{out})/\sqrt{\rm{Tr}(\rho_{th}\rho_{th})\rm{Tr}(\rho_{out}\rho_{out})}$, where the theoretic output state is $(|0\rangle+e^{-i\pi/4}|1\rangle)/\sqrt{2}$ with initial state in $(|0\rangle-i|1\rangle)/\sqrt{2}$. The maximal experimental unattenuated fidelity is found approach 0.996 for our scheme while the simulate one can reach 0.999 in absence of decoherence \cite{Supp}. As shown in Fig. 4(b) and (d), our gate is less sensitive to the detuning error than the conventional Holonomic gate. This can be attribute to our scheme remove the fixed pulse area limitation, thus further improve the robustness against control noises \cite{Zhusl2003,Zheng2004}.

\emph{Summary.}\textbf{--} We have experimentally demonstrated the single- and two-qubit gates for the TOUNHQC scheme. We characterize that our routine has a high gate fidelity by interleaved randomized benchmark. Moreover, we prove our two-qubit gate indeed more robust against certain control noises than the conventional NHQC gate by measuring the fidelity under these control imperfections. Therefore our scheme is a promising candidate to realize quantum state transfer and quantum gates in future.

This work was supported by the Key-Area Research and Development Program of GuangDong Province (Grant No. 2018B030326001), the NKRDP of China (Grant No. 2016YFA0301802), and the NSFC (Grants No. 11604103, No. 11474153, No. 91636218, No. 11890704, No. 61521001, No. 11875160 and No. U1801661), the Natural Science Foundation of Guangdong Province (Grant No.2017B030308003), the Guangdong Innovative and Entrepreneurial Research Team Program(Grant No.2016ZT06D348), the Science, Technology and Innovation Commission of Shenzhen Municipality (Grants No. JCYJ20170412152620376, No. JCYJ20170817105046702, and No. KYTDPT20181011104202253), the Economy, Trade and Information Commission of Shenzhen Municipality (Grant No.201901161512).

\end{document}